\documentclass[twocolumn,prl,showpacs]{revtex4}

\usepackage{graphicx}
\usepackage{rotating}
\usepackage{amsmath}
\usepackage{amsfonts}
\usepackage{amssymb}
\usepackage{enumerate}
\usepackage{longtable}
\usepackage{dcolumn}
\usepackage{bm}

\begin{document}

\title{Mott-Hubbard quantum criticality in paramagnetic CMR pyrochlores}

\author{L. Craco$^1$}
\author{C. I. Ventura$^2$}
\author{A. N. Yaresko$^3$}
\author{E. M\"uller-Hartmann$^1$}
\affiliation{$^1$ Institut f\"ur Theoretische Physik, Universit\"at zu K\"oln,
 Z\"ulpicher Stra{\ss}e 77, 50937 K\"oln, Germany}
\affiliation{$^2$Centro At\'omico Bariloche, 8400 - Bariloche, Argentina}
\affiliation{$^3$Max-Planck-Institut f\"ur Physik Komplexer Systeme,
N\"othnitzer Stra{\ss}e 38, 01187 Dresden, Germany}
\date{\rm\today}

\begin{abstract} 
We present a correlated {\it ab initio} description of the paramagnetic 
phase of Tl$_2$Mn$_2$O$_7$, employing a combined local density approximation 
(LDA) with multiorbital dynamical mean field theory (DMFT) treatment. 
We show that the insulating state observed in this colossal 
magnetoresistance (CMR) pyrochlore  is determined by strong Mn intra- and 
inter-orbital local electron-electron interactions. Hybridization effects 
are reinforced by the correlation-induced spectral weight transfer.  
Our result coincides with optical conductivity measurements, whose low 
energy features are remarkably accounted for by our theory. Based on this 
agreement, we study the disorder-driven insulator-metal transition of 
doped compounds, showing the proximity of Tl$_2$Mn$_2$O$_7$ to quantum 
phase transitions, in agreement with recent measurements.  
\end{abstract}


\maketitle

Among the manganese oxides exhibiting colossal magnetoresistance 
(CMR)~\cite{cmrreviews}, the family of pyrochlores represented by 
Tl$_2$Mn$_2$O$_7$,\cite{5,6,7} and the related compounds prepared 
by substitution of its different components~\cite{6,scdoping,bidoping} 
stands out. Though presenting similar coupling between magnetic 
and transport properties as the CMR perovskite manganites like  
${\rm La_{1-x}A_xMnO_3}$ (A= Ca,Sr,Ba), in the last nine years 
experiments have established many differences of their electronic 
properties\cite{5}, lattice behavior~\cite{5,7,shima-kei}, spin 
dynamics~\cite{lynn}, etc. In particular, transport and magnetism 
in CMR pyrochlores seem to be primarily related to different electronic 
orbitals~\cite{7,scdoping}  coupled by hybridization. Mechanisms 
like double exchange, involving the transfer of electrons between neighbor 
${\rm Mn^{3+}-Mn^{4+}}$ ions which favours ferromagnetic alignment of the 
Mn core spins~\cite{de}, and effects like Jahn-Teller distortions~\cite{jt} 
were early discarded for pyrochlore compounds. A series of alternative 
theoretical proposals were put forward and explored in connection with 
the experimental results~\cite{civba,littlewood,civmag}. The first 
microscopic model studied for Tl$_2$Mn$_2$O$_7$ was the intermediate 
valence model (IVM)~\cite{civba}, proposed to explore the 
suggestion~\cite{7} of the presence of a small effective internal doping 
of the type ${\rm Tl^{3+}_{2-x} Tl^{2+}_{x} Mn^{4+}_{2-x} Mn^{5+}_{x} O_7}$ 
(x$\sim$0.005). The presence of spin-dependent hybridization gaps (or 
pseudogaps), and the predicted temperature and magnetic field dependent 
changes of the electronic structure allowed description of the observed 
magnetotransport~\cite{civba}. In particular, the predicted evolution 
towards a gapped paramagnetic state above $T_c$  explained the drastic 
reduction of the number of carriers at $T_{c}$ in Hall data~\cite{5,7}. 
Recently, thermopower and optical conductivity were qualitatively described 
using the IVM\cite{foglio}. Majumdar and Littlewood~\cite{littlewood} 
explored the scenario of spin fluctuations around $T_{c}$ in the presence 
of very low carrier densities, first suggested in 
Refs.~\onlinecite{7,scdoping}. Considering Hund and superexchange 
couplings~\cite{littlewood}, they explained CMR in terms of spin polarons. 
Later, a generic effective model for Tl$_2$Mn$_2$O$_7$~\cite{civmag} 
was introduced to explore and compare various  
proposals~\cite{5,7,littlewood,mishra,dolores}. Interestingly, its 
electronic structure was shown~\cite{civmag}  to exhibit similar features 
to the IVM model~\cite{civba}, for appropriate parameters, and the 
experimental spin dynamics\cite{lynn} is  described~\cite{civacqua} 
if ferromagnetic superexchange is assumed\cite{dolores}. 
 
Tl$_{2}$Mn$_{2}$O$_{7}$ is cubic at room temperature, with an fcc 
($Fd\bar{3}m$) arrangement of corner-sharing Mn-O6 octahedra~\cite{5,7}. 
Its large magnetoresistance accompanies a ferromagnetic metal to 
paramagnetic insulator transition, with $T_c$ around $130$~K~\cite{5,6,7}. 
The magnetization below $T_c$ is believed to be determined by ferromagnetic 
superexchange coupling in the Mn$^{4+}$ sublattice~\cite{7,dolores}. Hall 
data indicate a very small electron-like carrier density~\cite{5} 
($\sim$ 0.001-0.005 $e$/f.u.), connected with the presence of extended 
Tl-$6s$ orbitals hybridizing with O-$2p$ and Mn-$t_{2g}$ states near the 
Fermi level~\cite{{bsingh,mishra,bshim}}.  Upon Bi-substitution on the 
Tl site, magnetoresistance increases drastically  and transport above $T_c$ 
is strongly modified~\cite{bidoping}. Notably, CMR is achieved at room 
temperatures, indicating the possibi\-li\-ty of technological applications 
of ${\rm Tl_{2-x}Bi_xMn_2O_7}$ and related compounds.

The electronic band structure of Tl$_2$Mn$_2$O$_7$ has been calculated 
in local density approximation (LDA) by various
groups~\cite{bsingh,mishra,bshim}. They obtained similar 
results for the ferromagnetic metallic ground state, characterizing it 
as a half-metal with minority-spin free-electron-like carriers. However,
no attention has been payed to the paramagnetic (PM) phase. In this work,  
we are presenting the first {\it ab initio} study of the paramagnetic 
phase of Tl$_2$Mn$_2$O$_7$. We found that LDA calculations predict a 
metallic paramagnetic state, in contrast to recent optical 
conductivity~\cite{condexp} and photoemission~\cite{photoexp} experiments 
clarifying its insulating nature. By including multiorbital correlations 
through a combination of LDA with dynamical field theory (DMFT), 
we characterize the paramagnetic phase as a Mott-Hubbard insulator, 
with a correlation-induced gap. We calculated optical conductivity 
contributions, and discuss our results in the context of recent 
experiments~\cite{condexp,photoexp} and effective model 
calculations~\cite{civba}. Finally, we focus on the insulator-metal 
transition induced by chemical substitution.

\begin{figure}[t]
\includegraphics[width=3.1in]{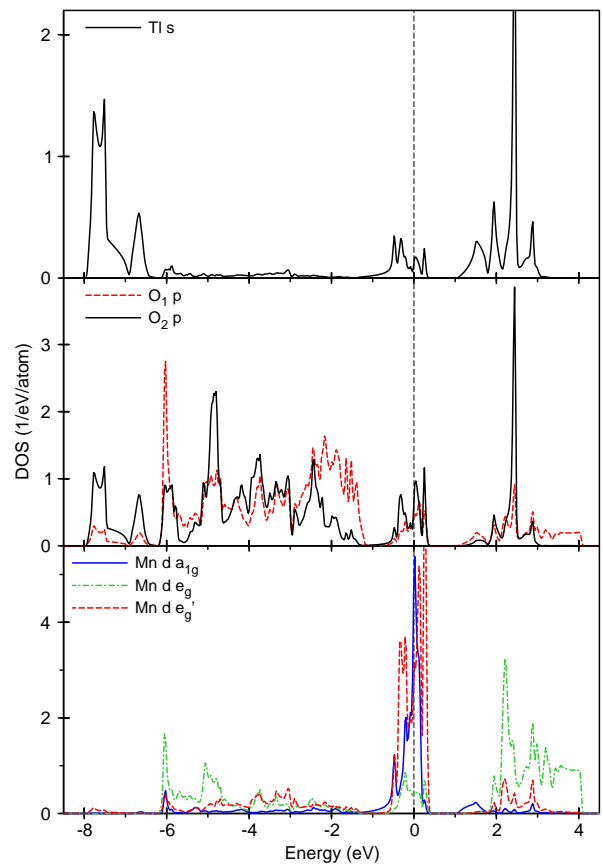}
\caption{(Color online) LDA band structure for paramagnetic 
${\rm Tl_2Mn_2O_7}$. O$_{1}$~(O$_{2}$) denotes Oxygen ions nearest to 
Mn (Tl).}
\label{lda-dos}
\end{figure}

LDA band structure calculations were performed for the experimental 
crystal structure of Tl$_2$Mn$_2$O$_7$~\cite{bshim} using the LMTO 
method~\cite{yaresko2} in the atomic sphere approximation. The overlap 
of atomic spheres was decreased by adding two sets of empty spheres in 
8$a$ and 32$e$ Wyckoff positions of the $Fd\bar{3}m$ space group. Our 
ferromagnetic phase results (not shown) agree with previous 
ones~\cite{bsingh,mishra,bshim}, while the electronic structure obtained 
for the paramagnetic phase is shown in Fig.~\ref{lda-dos}. The 
differentiated  hybridization between Tl and Mn with the two kinds of  
O atoms present is evident. Clearly, a metallic paramagnetic state 
is predicted by LDA, in contradiction with the insulating behavior 
recently established experimentally~\cite{condexp,photoexp}.   

Let us briefly outline the scheme of electronic structure calculation used. 
To include the real band structure and reliably treat the effect of 
correlations in the PM phase, as well as study metal-insulator transitions, 
we adopted the combined LDA+DMFT approach, which is becoming widely 
recognized as suitable for the realistic description of transition metal 
oxides. Previous similar applications of the technique include, e.g., 
the study of the insulator-metal transition in V$_2$O$_3$ and the 
ferromagnetic metallic state of CrO$_2$~\cite{paper3-4}. The multi-orbital 
many-body Hamiltonian considered for the LDA+DMFT study of 
Tl$_2$Mn$_2$O$_7$ is:

\begin{eqnarray}
\nonumber
H &=& \sum_{{\bf k} \alpha \beta \sigma}(\epsilon_{{\bf k} \alpha}
+\epsilon_{\alpha}^{0}\delta_{\alpha \beta}) 
c_{{\bf k} \alpha \sigma}^{\dag} c_{{\bf k} \beta \sigma} 
+ U \sum_{ia} n_{i \alpha \uparrow}n_{i \alpha \downarrow} \\ 
&+& U' \sum_{i \alpha \ne \beta} n_{i\alpha}n_{i\beta} 
- J_{H} \sum_{i \alpha \ne \beta} {\bf S}_{i\alpha}.{\bf S}_{i\beta} \;,
\label{eq1}
\end{eqnarray}
where $\alpha, \beta$ denote the three $t_{2g}$ orbitals. Due to the 
pyrochlore crystal field, the $t_{2g}$ sub-shell is split into an $a_{1g}$ 
singlet and an $e_{g}^{'}$ doublet. The size of the splitting 
$\Delta= \delta_{e_{g}^{'}} - \delta_{a_{1g}}$ ($\delta_\alpha$ being the 
center of gravity (c.g.) of the $\alpha$ band) within LDA is: 
$\Delta_{LDA}=0.037~eV$. To avoid double-counting of interactions included 
already in the LDA in average, $\epsilon_{\alpha}^{0}$ reads; 
$\epsilon_{\alpha}^{0}=\epsilon_{\alpha}-U(n_{\alpha \bar \sigma}-\frac{1}{2})
+\frac{J_H}{2} \sigma (n_{\alpha \sigma}-1)$, with $\epsilon_{\alpha}$ 
being the on-site energies of the $t_{2g}$ orbitals. The first term in 
Eq.~(\ref{eq1}) describes the one-electron part of the Hamiltonian, 
including details of the pyrochlore crystal structure of Tl$_2$Mn$_2$O$_7$ 
through the LDA bandstructure: $\epsilon_{{\bf k} \alpha}$. The next three 
terms account for local correlation effects in the $t_{2g}$ orbitals. 
Notice that we are keeping only the $t_{2g}$ orbitals since, from LDA,    
the density of states (DOS) for the $e_{g}$ bands is small near the Fermi 
level $E_{F} (\equiv 0)$, the $O-2p$ bands lie approximately $1.5~eV$ below 
$E_{F}$ and the Tl-$6s$ band about $2.5~eV$ above $E_{F}$. $U~(U')$ denotes 
the $t_{2g}$ intra- (inter-) orbital local Coulomb interaction. The last 
term in Eq.~(\ref{eq1}) describes the Hund's rule coupling: being J$_H$ 
poorly screened, we take it of the order of its atomic value in 
$Mn^{4+}$, $J_H=1~eV$. Rotational invariance fixes:   
$U' = U - 2 J_{H}$~\cite{rotinv}.

We solve Eq.~(\ref{eq1}) in $d=\infty$ using multi-orbital iterated 
perturbation theory (MO-IPT)~\cite{paper3-4}. Assuming no symmetry breaking 
in the spin/orbital sector, we have $G_{\alpha\beta\sigma\sigma'}(\omega)=
\delta_{\alpha\beta}\delta_{\sigma\sigma'}G_{\alpha\sigma}(\omega)$ and 
$\Sigma_{\alpha\beta\sigma\sigma'}(\omega)=\delta_{\alpha\beta}
\delta_{\sigma\sigma'}\Sigma_{\alpha\sigma}(\omega)$. The DMFT solution 
involves $(i)$ replacing the lattice model by a self-consistently embedded 
multi-orbital, asymmetric Anderson impurity model, and, $(ii)$ a  
selfconsistency condition requiring the local impurity Green's function (GF) 
to be equal to the local GF for the lattice. The calculation follows 
the philosophy of the one-orbital IPT, with the Green functions and 
self-energies being matrices in the orbital indices. The equations for 
the multi-orbital case are the same as used before~\cite{paper3-4}. They 
are solved selfconsistently with the LDA density of states as input, 
until convergence is achieved. 

\begin{figure}[t]
\includegraphics[width=\columnwidth]{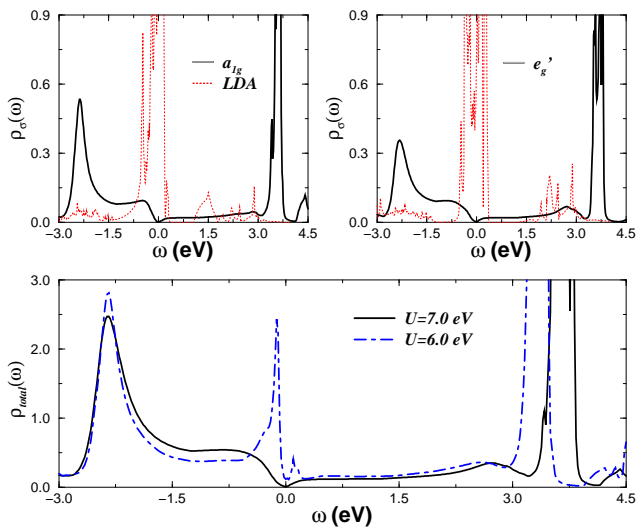}
\caption{(Color Online) LDA+DMFT (paramagnetic Tl$_{2}$Mn$_2$O$_7$): 
orbital-resolved (upper panels) and total (lower panel) Mn densities 
of states, for $U=6,7~eV$, $J_H=1~eV$.}
\label{fig2}
\end{figure}

We now present our LDA+DMFT results. In Fig.~\ref{fig2} we plot the 
$t_{2g}$ density of states of the Mn-ions. Using $U=7$~eV and $U'=5$~eV 
as intra- and interorbital correlation values, we obtain a clear 
Mott-Hubbard insulating state with an energy gap at the Fermi level. 
Compared to the LDA results of Fig.~\ref{lda-dos}, spectral weight has 
been transferred to the Hubbard satellites close to maxima of the Tl and 
O bands, therefore reinforcing hybridization effects. Thus, one may 
envisage an overall gapped (insulating) state resulting from rehybridization, 
in a scenario with common features to those predicted by the effective 
IVM for paramagnetic Tl$_2$Mn$_2$O$_7$ \cite{civba}. 
Recent photoemission experiments at room temperature~\cite{photoexp} 
confirmed the insulating nature of the paramagnetic phase. Finding a  
higher weight for  O-$2p$ at the upper edge of the valence band, in 
Ref.~\cite{photoexp} ${\rm Tl_2Mn_2O_7}$ is characterized as a 
charge-transfer insulator. Strong hybridization of  O-$2p$ and 
Mn-$t_{2g}$ in the valence band is reported, and a 2-3 eV difference 
between the centers of gravity of the respective bands is estimated: the 
Mn one lying mainly at binding energies above 3 eV~\cite{photoexp}. 
Interestingly, the  correlation-induced shift of Mn spectral weight we find 
with respect to the LDA, goes in the same direction: shifting the Mn 
valence band to higher binding energies. Furthermore, 
Ref.~\onlinecite{photoexp} characterizes ${\rm Tl_2Mn_2O_7}$ as 
close to a metal-insulator transition, in agreement 
with our results presented in Fig.~\ref{fig2}. 

We have employed our LDA+DMFT result for $U=7~eV$ to evaluate the $t_{2g}$ 
contribution to the optical conductivity in the paramagnetic phase. 
Here, we consider only the $t_{2g}$ intraband optical transitions: 
due to orthogonality of these orbitals, one would expect negligible
contributions from interband transitions~\cite{pavarini}. Within the 
$t_{2g}$-subshell the contributions to optical conductivity are calculated 
from: 

\begin{eqnarray}
\nonumber
\sigma (\omega) \propto  \sum_{\alpha \bf k} \int d\omega' 
A_{\alpha \bf k} (\omega') A_{\alpha \bf k} (\omega'+\omega) 
\frac{(f(\omega') - f(\omega+\omega')}{\omega}
\end{eqnarray}  
where $f(\omega)$ is the Fermi function and  
$A_\alpha ({\bf k},\omega)=\frac{1}{\pi} {\rm Im} [\omega - 
\Sigma_{\alpha}(\omega) - \epsilon_{\alpha \bf k} ]^{-1}$ is 
the spectral density. 

\begin{figure}[t]
\includegraphics[width=\columnwidth]{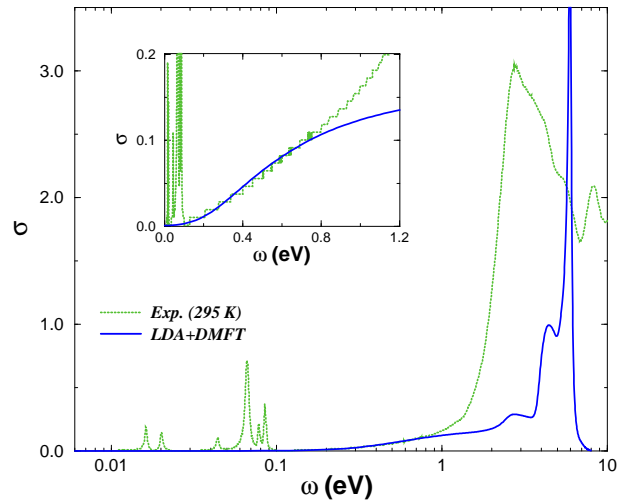}
\caption{(Color Online) Paramagnetic Tl$_{2}$Mn$_2$O$_7$: $t_{2g}$ optical
conductivity calculated with LDA+DMFT(U=7~eV), compared with expts.   
at T=295K~\protect\cite{condexp} Inset: Low-energy conductivity: LDA+DMFT 
vs. experiment at T=295K.}
\label{fig3}
\end{figure}

In Fig.~\ref{fig3} we show the calculated optical conductivity together 
with the experimental data at room temperature~\cite{condexp}. Apart from 
phonon related peaks (not included in our theory), a remarkable agreement 
is found at low energies (below 0.8 eV). The absence of optical response 
observed at very low energies evidences the insulating nature of the 
paramagnetic phase. For higher energies, a quantitative description of 
experiments would require to include the effect of inter-band 
charge-transfer excitations, from O-2p to Mn-3d bands, out of the scope 
of the present work.

\begin{figure}[t]
\includegraphics[width=\columnwidth]{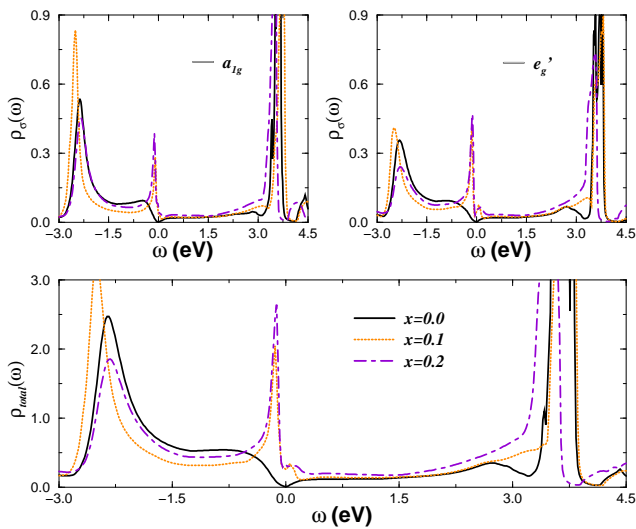}
\caption{(Color Online) Effect of disorder: 
orbital resolved (upper panels) and total (lower panel) DOS for 
Tl$_{2-x}$Bi$_x$Mn$_2$O$_7$. U= 7~eV, U'=5~eV and v= 2.5~eV.}
\label{fig4}
\end{figure}

Motivated by the remarkable agreement between our theory and the optical 
spectra, we now address the effect of disorder induced by chemical 
substitution~\cite{bidoping}. The problem is treated within the 
LDA+DMFT(IPT+CPA)~\cite{paper6} approach, which treats disorder exactly 
in high-dimensions and has been used to describe the doping-driven 
insulator-metal transition in ${\rm LaTiO_3}$. However, differently from 
Ref.~\onlinecite{paper6}, here we consider only the effect of 
substitutional disorder: without the introduction of extra holes or 
electrons in the system. [Notice that in a binary-alloy distribution for 
disorder (CPA), a fraction $x$ of the sites have an additional local 
potential $v$ for an electron (or hole) hopping onto that site.] By this, 
we aim to account for a host of experimental realizations in the limit of 
small impurity concentrations~\cite{doping}. Having in mind the possibility 
of technological applications  upon $Bi$-substitution~\cite{bidoping}, 
in our calculation we use a disorder potential $v= 2.5~eV$, corresponding 
to the energy difference between the Tl-$6s$ band and the Bi-$6p$ 
band~\cite{park}. Fig.~\ref{fig4} shows that introduction of a small 
amount of impurities metallizes the system, in agreement 
with~\cite{bidoping,doping}. Due to the pyrochlore structure, the 
Mott-Hubbard insulating state seems to be very unstable, allowing for 
first-order insulator-metal transitions upon small perturbations. 
According to our results, disorder or chemical pressure may drive  
Tl$_{2}$Mn$_2$O$_7$ into a bad metal fixed point, similar to that
obtained for $U=6$~eV (Fig.~\ref{fig2}). 

To conclude, we have presented the first {\it ab initio} study of the
paramagnetic phase of Tl$_{2}$Mn$_2$O$_7$. Through the inclusion of 
local multiorbital Coulomb interactions and using a combined LDA+DMFT 
approach, we describe the insulating nature of this phase.
Hybridization effects are reinforced by the spectral weight transfer due 
to strong correlations. The electronic structure obtained with $U=7$~eV 
allows us to provide not only a consistent description of the 
low-energy optical conductivity data, but also a strong support 
for our picture of disorder effects induced by chemical substitution. 
In agreement with chemical doping studies and recent photoemission findings, 
our results indicate that Tl$_{2}$Mn$_2$O$_7$ is very near to quantum 
phase instabilities, and illustrate the interplay of strong multi-orbital 
correlations with disorder and pyrochlore structure effects, 
thus calling for investigation with higher resolution spectroscopies, 
including inverse photoemission.

We benefited from discussions with D.I. Khomskii and M.S. Laad. 
C.I.V. thanks B. Alascio, J.A. Alonso, M. Foglio, M.N. Regueiro and
G. Zampieri for comments and references. LC was supported by the
SFB 608  of the Deutsche Forschungsgemeinschaft. C.I.V. is 
Investigador Cient\'{\i}fico of CONICET(Argentina), and thanks 
CONICET/DAAD 
for financial support as well as the Inst. f\"ur Theor. Physik, 
Univ. zu K\"oln, for hospitality.

\end{document}